\documentclass[fleqn,10pt]{wlscirep}
\usepackage[utf8]{inputenc}
\usepackage[T1]{fontenc}
\usepackage{upgreek}
\usepackage{color}

\title{An atomic Fabry-Perot interferometer using a pulsed interacting Bose-Einstein condensate}

\author[1]{P.~Manju}
\author[1]{K.~S.~Hardman}
\author[1]{P.~B.~Wigley}
\author[1]{J.~D.~Close}
\author[1]{N.~P.~Robins}
\author[1,*]{S.~S.~Szigeti}
\affil[1]{Atomlaser and Quantum Sensors Group, Department of Quantum Science, Research School of Physics, The Australian National University, Canberra 2601, Australia}

\affil[*]{stuart.szigeti@anu.edu.au}

\begin{abstract}
We numerically demonstrate atomic Fabry-Perot resonances for a pulsed interacting Bose-Einstein condensate (BEC) source transmitting through double Gaussian barriers. These resonances are observable for an experimentally-feasible parameter choice, which we determined using a previously-developed analytical model for a plane matter-wave incident on a double rectangular barrier system. Through numerical simulations using the non-polynomial Sch\"odinger equation -- an effective one-dimensional Gross-Pitaevskii equation -- we investigate the effect of atom number, scattering length, and BEC momentum width on the resonant transmission peaks. For $^{85}$Rb atomic sources with the current experimentally-achievable momentum width of $0.02 \hbar k_0$ [$k_0 = 2\pi/(780~\text{nm})$], we show that reasonably high contrast Fabry-Perot resonant transmission peaks can be observed using a) non-interacting BECs, b) interacting BECs of $5 \times 10^4$ atoms with $s$-wave scattering lengths $a_s=\pm 0.1a_0$ [$a_0$ is the Bohr radius], and c) interacting BECs of $10^3$ atoms with $a_s=\pm 1.0a_0$. 
Our theoretical investigation impacts any future experimental realization of an atomic Fabry-Perot interferometer with an ultracold atomic source.

\end{abstract}
\begin{document}
\newcommand{\sss}[1]{{\color{red}#1}}
\newcommand{\KandP}[1]{{\color{red}(#1)}}
\flushbottom
\maketitle

\thispagestyle{empty}

\section*{Introduction}

Understanding the different and complementary properties of atoms compared with photons has advanced both fundamental and applied physics. In direct analogy to optical systems, atomic matter-waves can be coherently focussed~\cite{gallatin_laser_1991}, reflected~\cite{balykin_reflection_1987}, diffracted~\cite{moskowitz_diffraction_1983} and interfered~\cite{becker_molecular_2011}. These basic atom-optical elements have been combined to construct more sophisticated analogue systems such as atomic waveguides~\cite{bongs_waveguide_2001}, atom lasers \cite{mewes_output_1997,robins_pumped_2008} and atom interferometers~\cite{berman_contributors_1997,keith_interferometer_1991}. Atomic properties such as mass, tunable dispersion and differing degrees of freedom make these analogue systems versatile measurement tools. Atom interferometers, for example, have enabled state-of-the-art measurements of the fine structure constant~\cite{Bouchendira:2011, parker_measurement_2018} and inertial fields such as gravity \cite{Peters:2001,altin_precision_2013,freier_mobile_2016} and rotations~\cite{stockton_absolute_2011,Savoie:2018}.

In this paper, we consider the atomic analogue of a Fabry-Perot interferometer. Optical Fabry-Perot interferometry is used for many fundamental scientific and industrial applications, including linewidth measurements of continuous wave (CW) and pulsed lasers~\cite{xue_pulsed_2016}, laser phase and frequency stabilisation~\cite{drever_laser_1983} and precision sensing~\cite{taylor_principles_1998,islam_chronology_2014}. An atomic Fabry-Perot interferometer could offer new sensing capabilities by exploiting the atomic mass and tunable dispersion. Furthermore, the analogous mirrors, formed using optical potentials, allow for real-time and versatile control of the system. Previous theoretical work has investigated the resonance properties of Fabry-Perot interferometry using matter-waves~\cite{dutt_smooth_2010} and their potential use in velocity selection~\cite{ruschhaupt_velocity_2005,damon_reduction_2014,Valagiannopoulos:2019} and in the identification of bosonic and fermionic isotopes of an element~\cite{lee_optical_2005,wilkens_fabry-perot_1993}. In order to fully exploit the benefits of this atomic analogue, the transmission characteristics of an atomic Fabry-Perot interferometer in an experimentally-realisable regime must be understood.

Much like the optical Fabry-Perot interferometer, the atomic analogue requires a narrow linewidth source to fully exploit the interference effects of the system. The properties of an optical laser make it a superior source for a Fabry-Perot interferometer compared with a broadband light source. Bose-Einstein condensates (BECs) display many properties analogous to a laser, including high coherence and narrow momentum width, which have proven advantageous for precision atom interferometry~\cite{Debs:2011, Szigeti:2012, hardman_simultaneous_2016}. Although many theoretical proposals have shown that atomic Fabry-Perot interferometers can be developed using CW atom laser beams constructed from interacting and non-interacting BEC sources \cite{rapedius_barrier_2008,carusotto_nonlinear_2001,damon_reduction_2014,paul_nonlinear_2005,paul_nonlinear_2007,ernst_transport_2010}, a true CW atom laser has yet to be experimentally realised~\cite{Robins:2013}. In contrast, a pulsed atom laser, formed by releasing and propagating a BEC, is readily achievable in ultracold-atom laboratories. 

This paper investigates the properties of an atomic Fabry-Perot interferometer in an experimentally-realisable parameter regime~\cite{everitt_observation_2017,kovachy_matter_2015}. We use a simple analytic model~\cite{xiao_resonant_2015} for rectangular barrier potentials to study the dependence of cavity finesse and transmission coefficient on barrier width, barrier height, cavity length, initial momentum of the atoms, and the momentum width of the atomic cloud. This allows us to identify a parameter regime where high contrast, narrow peaks are observable in the transmission spectrum. Using these established parameters, we numerically simulate the propagation of a pulsed $^{85}$Rb BEC through a cavity formed via double Gaussian barriers. $^{85}$Rb is ideal for this study since its inter-atomic interactions are tunable via a Feshbach resonance~\cite{roberts_resonant_1998,kuhn_bose-condensed_2014,mcdonald_bright_2014}. We investigate the resonant transmission process for an interacting and non-interacting BEC and study the effect of the condensate's momentum width and inter-atomic interactions on resonant transport through the double barrier system. Using the experimentally feasible parameters we have determined, we demonstrate resonant transmission, requiring a momentum width achievable using delta-kick cooling \cite{ammann_delta_1997,kovachy_matter_2015, mcdonald_$80ensuremathhbark$_2013}. We show that the transmission coefficient and finesse can be improved by reducing the cloud's initial momentum width, allowing for high contrast resonant transmission in the weakly-interacting regime.

\begin{figure}[t]
\centering
\includegraphics[width=1\linewidth]{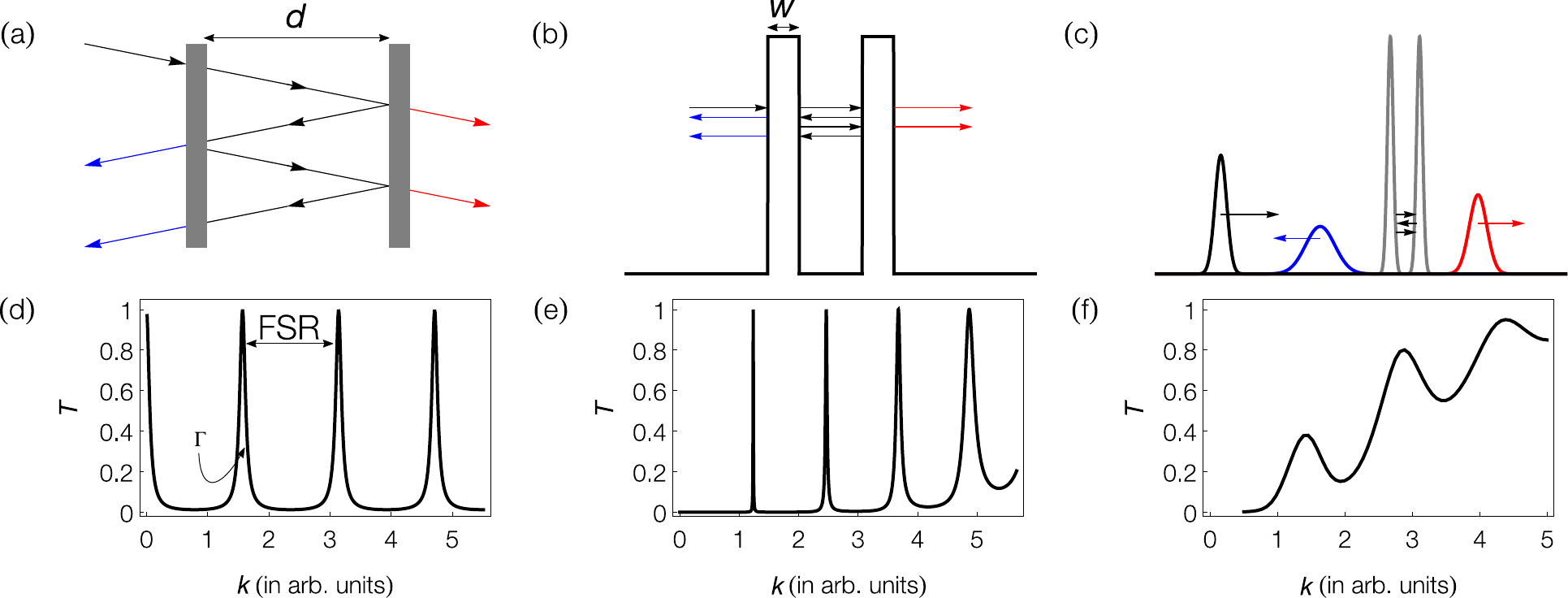}
\caption{Schematic diagrams of Fabry-Perot interferometers and their respective transmission spectra. The top row depicts the transmission and reflection of (a) narrow band light through an optical Fabry-Perot interferometer, (b) a narrow-band matter-wave through a rectangular double barrier, and (c) a broad-band Gaussian wavepacket (i.e. a non-interacting BEC with finite momentum width) through double Gaussian barriers. Outside the cavity, the black arrow represents the incident wave, whilst blue and red denote the reflected and transmitted parts of the wave, respectively. (d), (e) and (f) show corresponding transmission resonance peaks. Here $d$ is the cavity length, $w$ is the barrier width, $T$ is the transmission coefficient, FSR is the free spectral range of the resonance spectrum, and $\Gamma$ is the cavity linewidth. }
\label{fig:Schematic}
\end{figure}

\section*{Fabry-Perot Interferometry}\label{Analytics} 
In order to build an understanding of the parameter dependencies, we initially consider an idealised optical Fabry-Perot cavity, extending this to an analytic model describing the atomic analogue. Using this analytic model in tandem with known experimental limitations, we determine a feasible parameter regime for realising an atomic Fabry Perot interferometer. These parameters are then used to simulate the more complex model involving barriers described by Gaussian potentials and a finite momentum width BEC, including inter-atomic interactions. This progression is illustrated in Fig.~\ref{fig:Schematic}. 

An ideal optical Fabry-Perot interferometer is made of two parallel mirrors separated by a distance, \textit{d} (cavity length), as shown in Fig.~\ref{fig:Schematic}(a). The light entering the cavity undergoes multiple reflections from the mirrors and interferes with itself. Constructive interference enhances the light inside the cavity, leading to resonant transmission out of the cavity. Here we consider the ideal case where the mirror reflectivity is independent of the wavelength of light. Resonant transmission peaks, illustrated in Fig.~\ref{fig:Schematic}(d), are obtained by scanning the wave number ($k$) of the incident light. These peaks have a linewidth~\cite{vaughan_fabry-perot_1989}
\begin{align}
    \Gamma &= \frac{c}{2\pi d}\frac{1-\sqrt{R_1 R_2}}{\left(R_1 R_2\right)^{1/4}}\,,
\end{align}
and are separated by the free spectral range 
\begin{align}
    \text{FSR} &= \frac{c}{2 d},
\end{align}
where $R_1$ and $R_2$ describe the reflectivity of each mirror forming the cavity, and $c$ is the velocity of light in vacuum.
These quantities, along with the finesse
\begin{align}
    \mathcal{F} &= \frac{\text{FSR}}{\Gamma}=\frac{\pi (R_1R_2)^{1/4}}{1-\sqrt{R_1R_2}},
\end{align}
are the figures of merit for an optical Fabry-Perot interferometer~\cite{vaughan_fabry-perot_1989}. 

\begin{figure}[t]
\centering
\includegraphics[width=0.9\linewidth]{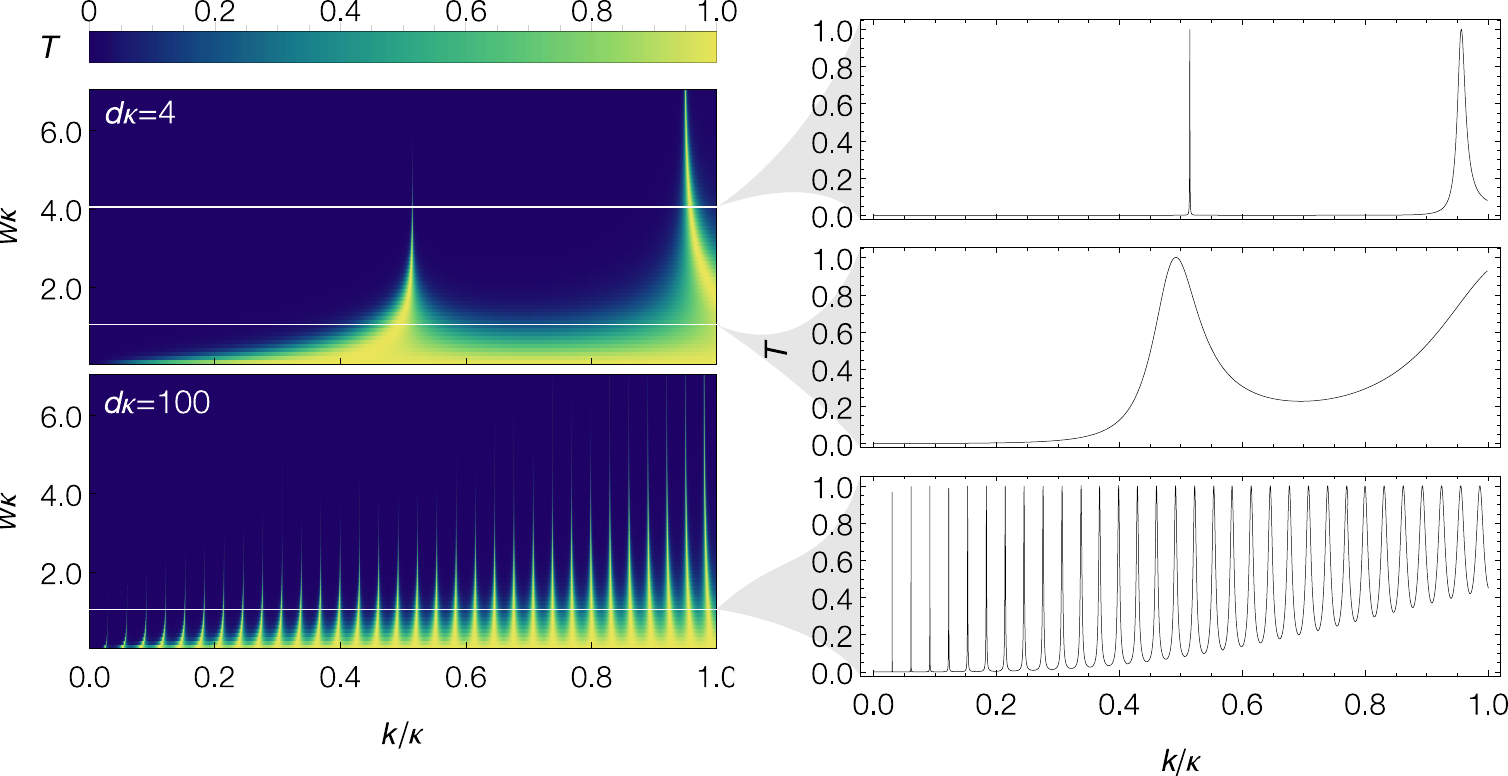}
\caption{Transmission coefficient of a narrow-band beam of particles passing through a double rectangular barrier system [Eq.~(\ref{eq:dimensionlessTransmission}) - see also Fig.~\ref{fig:Schematic}(b) and (d)] for (top left) $d\kappa=4$ and (bottom left) $d\kappa=100$. The three plots on the right-hand side show cross sections from the left plot at (top) $w\kappa=4, d\kappa = 4$, (middle) $w\kappa = 1, d\kappa = 4$, and (bottom) $w\kappa = 1, d\kappa = 100$. For a fixed barrier height (fixed $\kappa$), the cavity linewidth decreases with increasing barrier width and/or cavity length, whereas the FSR decreases with increasing cavity length. Here $w, d, k$ and $\kappa$ are the barrier width, cavity length, and wave vectors of the particles and barriers, respectively.
} 
\label{fig:analytics}
\end{figure}
 
In an atomic Fabry-Perot interferometer, the incoming light is replaced by atomic matter-waves and the mirrors are replaced by laser-induced potential barriers. The atomic system provides key differences including atomic mass, inter-atomic interactions, and mirrors where the key parameters can be tuned. This results in changes to the transmission characteristics of the atomic system compared with that of the optical system. This is qualitatively illustrated by Fig.~\ref{fig:Schematic}(e) and (f). In contrast to the ideal optical Fabry-Perot interferometer, the atomic system displays changes in the FSR and $\Gamma$ as the wave number of the incoming particles changes. To analyse the characteristics of a matter-wave Fabry-Perot interferometer, we use a previously developed analytical model~\cite{xiao_resonant_2015}. This model assumes a plane wave of non-interacting particles, with energy $E$, transmitting through two symmetric rectangular barriers of width $w$, height $V_0$, and separation distance $d$ [see Fig.~\ref{fig:Schematic}(b)]. 

The transmission coefficient, $T$, can be described using the dimensionless parameters $d \kappa$, $w \kappa$, $k/\kappa$, where $k = \sqrt{2 m E/\hbar}$ and $\kappa = \sqrt{2 m V_0/\hbar}$ are the wave vectors of particles and barriers, respectively. Specifically,
\begin{equation} \label{eq:dimensionlessTransmission}
    T(d \kappa, w \kappa, k/\kappa) = \frac{1}{1 + \frac{4}{\pi^2}[\mathcal{F}(w \kappa, k/\kappa)]^2 \sin^2[d \kappa \times k/\kappa + \phi(w \kappa, k/\kappa)]},
\end{equation}
where
\begin{align}
    \mathcal{F}(w \kappa, k/\kappa) &= \frac{\pi \sqrt{R(w \kappa, k/\kappa)}}{1-\sqrt{R(w \kappa, k/\kappa)}}, \label{Eqn:F} \\
    R(w \kappa, k/\kappa)   &= \frac{[M_+(k/\kappa)]^2}{[M_-(k/\kappa)]^2 + \coth^2\left( w \kappa \sqrt{1 - (k/\kappa)^2}\right)}, \\
    M_\pm(k/\kappa) &= \frac{1}{2}\left(\frac{\sqrt{1-(k/\kappa)^2}}{(k/\kappa)} \pm \frac{(k/\kappa)}{\sqrt{1-(k/\kappa)^2}}\right), \\
    \phi(w \kappa, k/\kappa) &= \frac{\pi}{2} - \tan^{-1}\left[ M_-(k/\kappa) \tanh\left( w \kappa \sqrt{1 - (k/\kappa)^2}\right) \right].
\end{align}
Here $\mathcal{F}$ and $R$ are interpreted as the cavity's finesse and mirror (rectangular barrier) reflectivity, respectively.
The relationship of the transmission coefficient to $w\kappa$ and $k/\kappa$ is illustrated for two different values of $d\kappa$ in Fig.~\ref{fig:analytics}. The three plots on the right-hand side show cross sections from the left plot for $d\kappa=4$ with $w\kappa=4$ (top) and $w\kappa=1$ (middle) in addition to $d\kappa=100$ with $w\kappa=1$ (bottom). The first two cross sections show that the cavity linewidth decreases with increasing barrier width, for a fixed barrier height and cavity length. The second and third cross sections show that both FSR and linewidth decrease with increasing cavity length, for a fixed barrier height and width. In contrast to the ideal optical Fabry-Perot interferometer, the linewidth increases with increasing $k$ for fixed barrier height and barrier width. This behaviour is due to the strong wavelength-dependence of the mirror reflectivity in the atomic Fabry-Perot interferometer. Similar effects would be seen when considering an optical cavity with strong wavelength-dependent reflectivities.

\begin{figure}[t]
\centering
\includegraphics[width=0.7\linewidth]{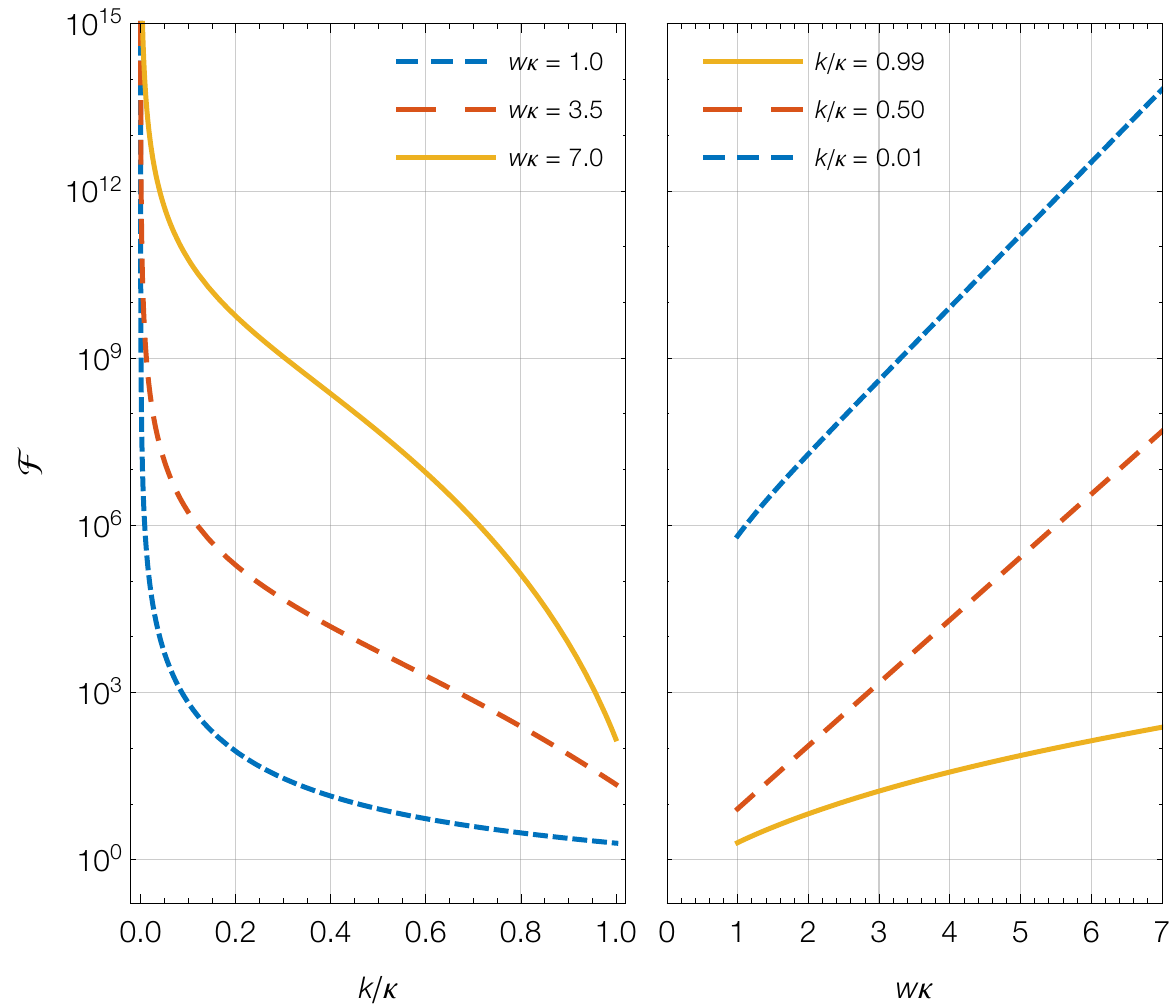}
\caption{For a beam of particles transmitting through a double rectangular barrier system, finesse is plotted as a function of dimensionless particle wave vector ($k/\kappa$) and barrier width $w\kappa$, where $\kappa = \sqrt{2m V_0} / \hbar$ is the wave vector corresponding to the barrier. For a fixed barrier height, finesse decreases with increasing momentum of the particle, but increases with increasing barrier width.}
\label{fig:finesse}
\end{figure}

Now we consider the finesse of the cavity, which is given by Eq.~(\ref{Eqn:F}). The finesse only depends upon the dimensionless parameters $w \kappa$ and $k/\kappa$. As shown in Fig.~\ref{fig:finesse}, the finesse decreases with increasing $k/\kappa$ and increases with increasing $w\kappa$. Therefore, for a fixed barrier height, an increase in particle momentum causes a decrease in finesse, whereas an increase in barrier width causes an increase in finesse. Again, this behaviour is different to the ideal optical Fabry-Perot interferometer, and is due to the mirror reflectivity's dependence on barrier height, barrier width, and the energy of the incoming atoms.

The above analysis reveals the parameter regimes needed to make an atomic Fabry-Perot interferometer with desirable qualities such as high finesse and narrow linewidth. However, we do not have complete freedom in our parameter choice, as the limitations of current cold-atom technology impose additional constraints. We discuss these additional constraints below, and combine them with the results of the above analytic model to determine an experimentally-feasible parameter regime for realising an atomic Fabry-Perot interferometer. The operation of an atomic Fabry-Perot interferometer depends sensitively on the following parameters:
\begin{itemize}
    \item \textbf{Momentum width of the atomic cloud.}
       The above analytical study assumes a plane matter-wave with infinitely narrow momentum width. In reality, all atomic clouds have finite momentum widths. In order to observe Fabry-Perot resonances, ideally the momentum spread of the atomic source needs to be much less than the FSR and linewidth of the cavity. To date, the smallest experimentally-achieved momentum width for an atomic cloud is $0.02 \hbar k_0$~\cite{kovachy_matter_2015}, which was obtained by delta-kick cooling a rubidium BEC (here $k_0= 2\pi / (780~\text{nm}) = 8.06\times 10^6 \text{m}^{-1}$ is the wave vector of the light used to impart momentum through Bragg spectroscopy on the $^{85}$Rb $D_2$ transition~\cite{adam_steck_rubidium_2013,stenger_bragg_1999,ernst_probing_2010}). Hence, we aim to select parameters that give a cavity linewidth and FSR larger than this value.   
\item \textbf{Barrier width.} 
    Although narrow resonances and a high finesse are desirable, the lower bound on experimentally-achievable BEC momentum width requires us to operate in a regime where the cavity linewidth is relatively broad. We find that selecting $w \kappa = 1$ is a good compromise, since it gives resonance peaks that are wide enough to be observable yet still narrow enough to be potentially useful. Additionally, the minimum achievable barrier width is restricted by the diffraction limit of the laser, which for our system is on the order of 1$\upmu$m. We fix the barrier width to this minimum, i.e. $w = 1~\upmu$m; a significantly larger choice for $w$ would require a smaller $\kappa$, resulting in a cavity spectrum that could only be observed by scanning unachievably small values of $k$.
    
    \item \textbf{Barrier height.} 
        Fixing $w$ and $w\kappa$ completely determines the barrier height. Explicitly, $\kappa= 1\upmu \text{m}^{-1}=0.12414 k_{0}$, corresponding to a barrier height of $V_0= \hbar^2 \kappa^2/(2m)=3.944\times 10^{-32}$ J.
    \item \textbf{Cavity length.}
        The finesse of the cavity does not depend on the cavity length. However, in order to reduce the overlap between the two laser-induced barriers experimentally, we need a cavity length that is larger than the barrier width. Additionally, momentum width considerations require a regime where the cavity linewidth is relatively broad (see above). Figure~\ref{fig:analytics} shows that as $d\kappa$ decreases, both FSR and linewidth increase. Therefore, in order to observe at least two broad linewidth transmission peaks, we choose $d\kappa=4$. This gives a cavity length of $d= 4 \upmu \text{m}$, which is larger than the chosen barrier width ($d=4 w$). 
    \item \textbf{Momentum imparted to the atoms.}
        Finally, in order to observe resonance peaks, we must be able to scan the incident energy of the atoms. This can be done by imparting momentum to the atoms through, for example, Bragg transitions~\cite{stenger_bragg_1999,altin_precision_2013}. For the above parameter choices, at least two peaks are observable by scanning $k/\kappa$ from 0.01 to 1.2 (see Fig.~\ref{fig:analytics}). This corresponds to a $k$ range of $k = 0.0013k_0$ to $k = 0.15 k_0$.
   \end{itemize} 
Based on the above experimentally-feasible parameter choice, our simple analytic model predicts that resonant peaks should be observable with $\text{FSR}=0.0687 \hbar k_0$ (which is greater than the cloud momentum width, $0.02 \hbar k_0$) and linewidth $\Gamma=0.0097 \hbar k_0$. This linewidth is about a factor of two smaller than the momentum width of the cloud, which is not ideal. Nevertheless, as we show below with more detailed theoretical modelling, Fabry-Perot resonances are observable in this regime. Indeed, the experimental system includes much more complexity than the simple analytical model considered above. For example, the barriers are created experimentally using blue-detuned lasers, which are more accurately modelled as Gaussian barriers, in contrast to the rectangular barriers used in the analytical model. Furthermore, BECs are finite momentum width sources that typically have non-negligible inter-atomic interactions. These inter-atomic interactions couple different momentum components of the cloud and also have a non-trivial effect on the transmission dynamics~\cite{Salasnich:2001, Carr:2005, manju_quantum_2018, Wales:2020}. Therefore, although the analytic study can guide our parameter choice, it cannot provide detailed modelling of an interacting BEC's transmission dynamics through double Gaussian barriers. This demands a numerical investigation. 

\section*{Theoretical model for numerical simulation}\label{modelsystem}
The mean-field dynamics of a weakly-interacting BEC in a quasi-1D geometry (e.g. a waveguide potential with tight radial confinement) are well-described by the non-polynomial Schr\"odinger equation (NPSE)~\cite{Salasnich:2002,Wigley:2017,everitt_observation_2017,Carli:2019}. This is an effective 1D model of the Gross-Pitaevskii equation that incorporates spatial and temporal variations in the BEC's width in the tight radial direction (see Appendix). It assumes a cylindrically-symmetric tight harmonic radial confinement of frequency $\omega_\perp$. Specifically,
\begin{align} \label{Eq.1DGPE}
   i\hbar \frac{\partial\psi(z,t)}{\partial t}&=\left[-\frac{\hbar^2}{2m}\frac{\partial^2}{\partial z^2} + V(z) + \frac{g}{2 \pi a_\perp^2}\frac{|\psi(z,t)|^2}{\sigma(z,t)^2} - i\frac{\hbar K_3}{6 \pi^2 a_\perp^4} \frac{|\psi(z,t)|^4}{\sigma(z,t)^4} + \frac{\hbar \omega_\perp}{2}\left( \frac{1}{\sigma(z,t)^2} + \sigma(z,t)^2\right)\right]\psi(z,t), 
\end{align}
where $\sigma(z,t)^2 = \sqrt{1 + 2 a_s |\psi(z,t)|^2}$, $\psi(z,t)$ is the effective 1D macroscopic condensate wave function normalised to the total particle number, $N(t)=\int dz \, |\psi(z,t)|^2 $, $V(z)$ is the external potential (a double Gaussian barrier potential during evolution), $m$ is the atomic mass, $a_\perp = \sqrt{\hbar / (m \omega_\perp)}$, $g = 4 \pi \hbar^2 a_s / m$ is the two-body interaction strength, determined via the $s$-wave scattering length, $a_s$, and $K_3$ is the three-body recombination loss rate coefficient. In the case of an atomic Fabry-Perot interferometer, the three-body recombination loss rate can become significant due to the high density formed by the multiple reflections of the atoms between barriers. We set the three-body recombination loss rate coefficient to $K_3=4 \times 10^{-41} \text{m}^6/\text{s}$, which is the value determined from our previous experiments with $^{85}$Rb BEC~\cite{altin_collapse_2011, everitt_observation_2017}. All our simulations of Eq.~(\ref{Eq.1DGPE}) were performed for condensates of $^{85}$Rb atoms using the open-source software package XMDS2~\cite{Dennis:2012} with an adaptive 4th-5th order Runge-Kutta interaction picture algorithm.

The initial wave function for the simulations is a Gaussian,
\begin{align}
    \psi_0(z)= \frac{\sqrt{N}}{(\pi \sigma_c^2)^{1/4}} e^{-(z-z_0)^2/(2\sigma_c^2)}e^{i k z},
\end{align}
where $z_0$ is the initial position of the atomic cloud, $\sigma_c/\sqrt{2}$ is the standard deviation of the density profile, corresponding to 
a $k$-space density standard deviation of $1/(\sqrt{2}\sigma_c)$ (full width at half maximum (FWHM) of $\Delta k = 2 \sqrt{\ln 2}/\sigma_c)$, and $\hbar k$ is the condensate's initial mean momentum. A Gaussian wave packet of this functional form could be engineered experimentally by delta-kick cooling~\cite{ammann_delta_1997,kovachy_matter_2015,mcdonald_$80ensuremathhbark$_2013} the cloud after turning off the axial confinement, and then imparting a momentum kick $\hbar k$ to the atoms via a shallow angle Bragg transition~\cite{stenger_bragg_1999,ernst_probing_2010,hardman_simultaneous_2016}.

The laser barriers that form the Fabry-Perot cavity are  modelled as two Gaussian potentials: 
\begin{align}
    V(z)=V_0\big[ e^{{-(z-z_{01})^2}/(2\sigma_b^2)}+e^{{-(z-z_{02})^2}/(2\sigma_b^2)}\big],
\end{align}
where $\sigma_b$ is the standard deviation of each barrier and $z_{01}$ and $z_{02}$ are the position of first and second barriers, respectively. We choose $z_{01}=z_0+3(\sigma_c+\sigma_b)+15 \mu$m and $z_{02}=z_{01}+3\sigma_b+d+3\sigma_b$ so that introducing the barrier potentials does not perturb the initial atomic cloud. Guided by our previous analytic analysis, we choose cavity length $d = 4\mu$m, barrier height $V_0 = 3.944\times 10^{-32}$ J, and barrier width $\sigma_b = 1 \mu\textrm{m} / \sqrt{2\pi}$ (this gives a Gaussian barrier with equal area to a square barrier of height $V_0$ and width $1 \mu$m).

\begin{figure}[t]
\centering
\includegraphics[width=0.8\linewidth]{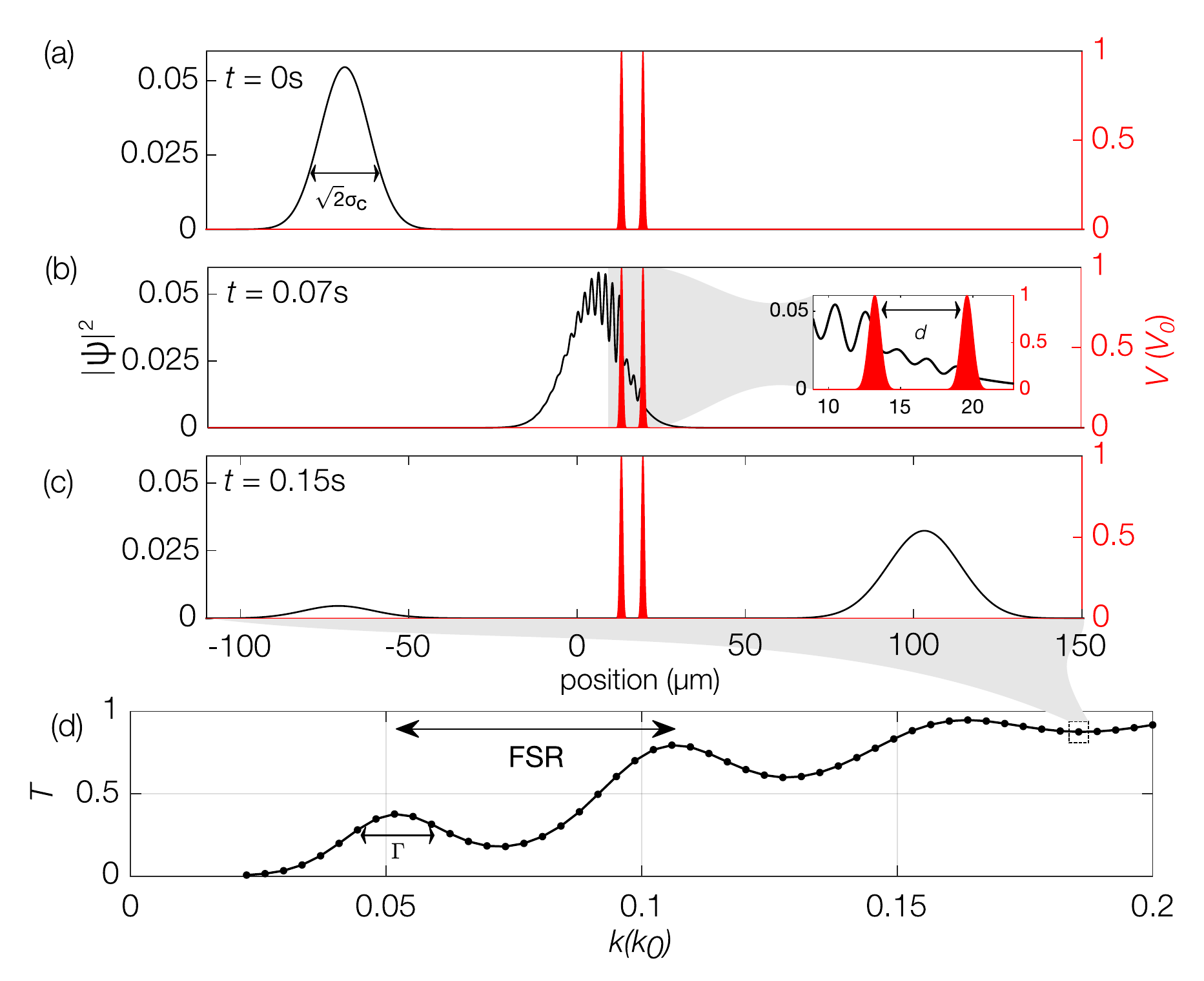}
\caption{(a), (b) and (c) show the propagation of a non-interacting BEC having mean momentum $\hbar k = 0.182 \hbar k_0$ and FWHM momentum width $\hbar \Delta k = 0.02 \hbar k_0$ through two Gaussian barriers, in three snapshots in time, simulated using the 1D Schr\"odinger equation (i.e. NPSE Eq.~(\ref{Eq.1DGPE}) with $g = 0$ and $K_3 = 0$). The red curves indicate the Gaussian barriers and the black curves represent the density profile of the Gaussian cloud. (a) The BEC starts on the left side of the barriers, (b) propagates towards the barriers, and enters the cavity. The dynamics around the cavity region is expanded and illustrated in the inset plot. (c) Some parts of the cloud are transmitted through the barriers, whereas other parts are reflected, depending on the momentum of the cloud. (d) The transmission coefficient as a function of the mean momentum of the wave packet [here $k_0=2\pi/(780$nm)].}
\label{fig:modelsystem}
\end{figure}

Our simulations allow us to determine the number of atoms transmitted ($N_T$) and reflected ($N_R$) through the double barrier system via
\begin{align}
    N_T  &= \int_{z_T}^{\infty} dz \, |\psi(z,t_\textrm{end})|^2, \\
    N_R  &= \int_{-\infty}^{z_\text{R}} dz \, |\psi(z,t_\textrm{end})|^2,
\end{align}
where $z > z_T=(z_{02}+3\sigma_\text{b})$ and $z<z_\text{R}=(z_{01}-3\sigma_\text{b})$ are the transmitted and reflected regions, respectively. The stopping time, $t_\text{end}$, for the simulation is chosen such that there are no atoms left in the cavity (i.e. $N-N_T-N_R < 1$) and both $N_T$ and $N_R$ have reached a constant value (more precisely, do not change by more than 0.1 in a given time step)\footnote{For the non-interacting case $g = 0$ and $K_3 = 0$, $N$ is simply a normalisation factor that no longer influences the dynamics. In this case, $t_\text{end}$ is chosen such that $(N_T+N_R)/N < 10^{-5}$ and both $N_T/N$ and $N_R/N$ do not change by more than $10^{-6}$ in a given time step.}. We investigate the resonant transmission by computing the total transmission coefficient
\begin{align}
    T &= \frac{N_T}{N_T+N_R}.
\end{align}

\begin{figure}[t]
\centering
\includegraphics[width=0.7\linewidth]{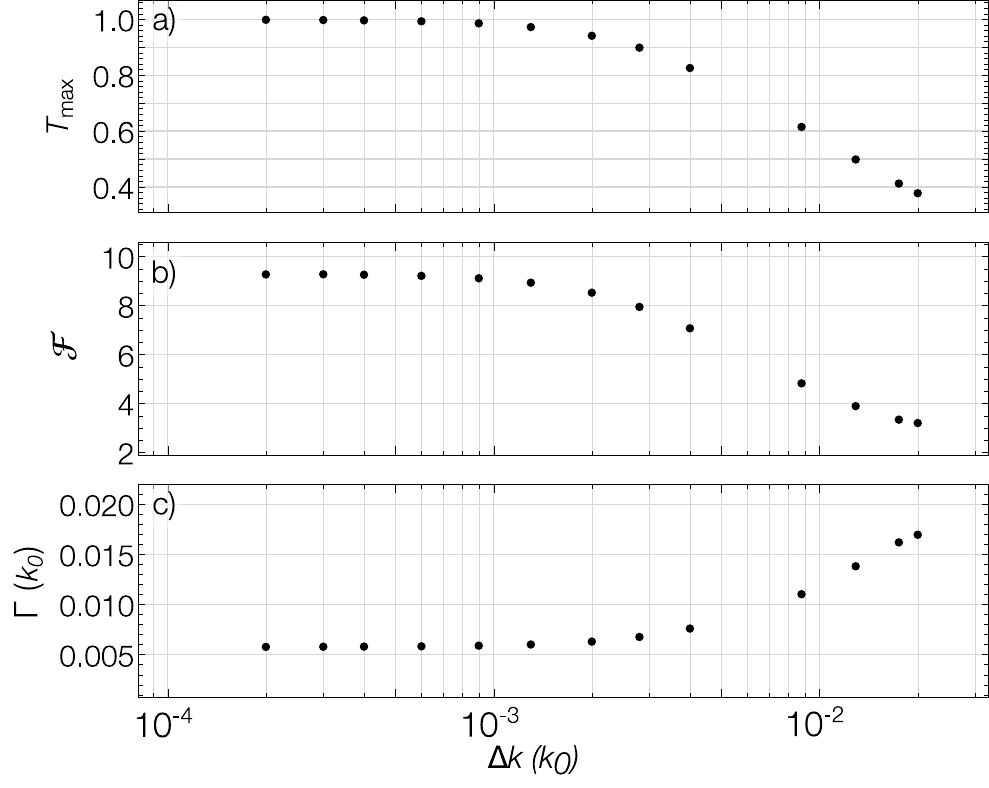}
\caption{The influence of a non-interacting BEC's momentum width on the (a) height, (b) finesse, and (c) linewidth of the first resonance peak. The FWHM of the initial momentum distribution, $\hbar \Delta k = 2 \sqrt{\ln 2} \hbar / \sigma_c$, and linewidth are given in units of $k_0=2\pi/(780$nm). Peak height and finesse increase and $\Gamma$ decreases with decreasing $\Delta k$.}
\label{fig:Varyingkwidthas_0}
\end{figure}
\section*{Analysis of resonant transmission} 
\subsection*{Non-interacting case}
Using the parameters determined above, we first simulate the resonant transmission of a non-interacting BEC ($g = 0$ and $K_3 = 0$) passing through the double Gaussian barrier system described previously. We choose an initial cloud with a FWHM momentum width close to the smallest experimentally-realised value of $\Delta k=0.02 k_0$~\cite{kovachy_matter_2015}, which corresponds to a spatial width of $2\sigma_c\approx 21 \upmu$m. The propagation of this non-interacting BEC after momentum kick $\hbar k =0.1855 \hbar k_0$ is schematically shown in Fig.~\ref{fig:modelsystem}(a), (b) and (c), where (a), (b) and (c) correspond to three snap shots in time: prior to interaction ($t=0$s), during interaction ($t=0.07$s), and after interaction ($t=0.15$s) with the barrier, respectively. 
The interference caused from the overlapping incident and reflected cloud components is clearly seen in Fig.~\ref{fig:modelsystem}(b). As expected, this behaviour is present both after the initial reflection, outside of the cavity, and through multiple reflections inside the cavity. Resonant transmission is observed for the $k$ values which are resonant with the cavity. The transmission coefficient as a function of momentum kick given to the cloud is plotted in Fig.~\ref{fig:modelsystem}(d). This confirms that Fabry-Perot resonances can indeed be observed for our parameter choice.

\begin{figure}[t]
\centering
\includegraphics[width=0.8\linewidth]{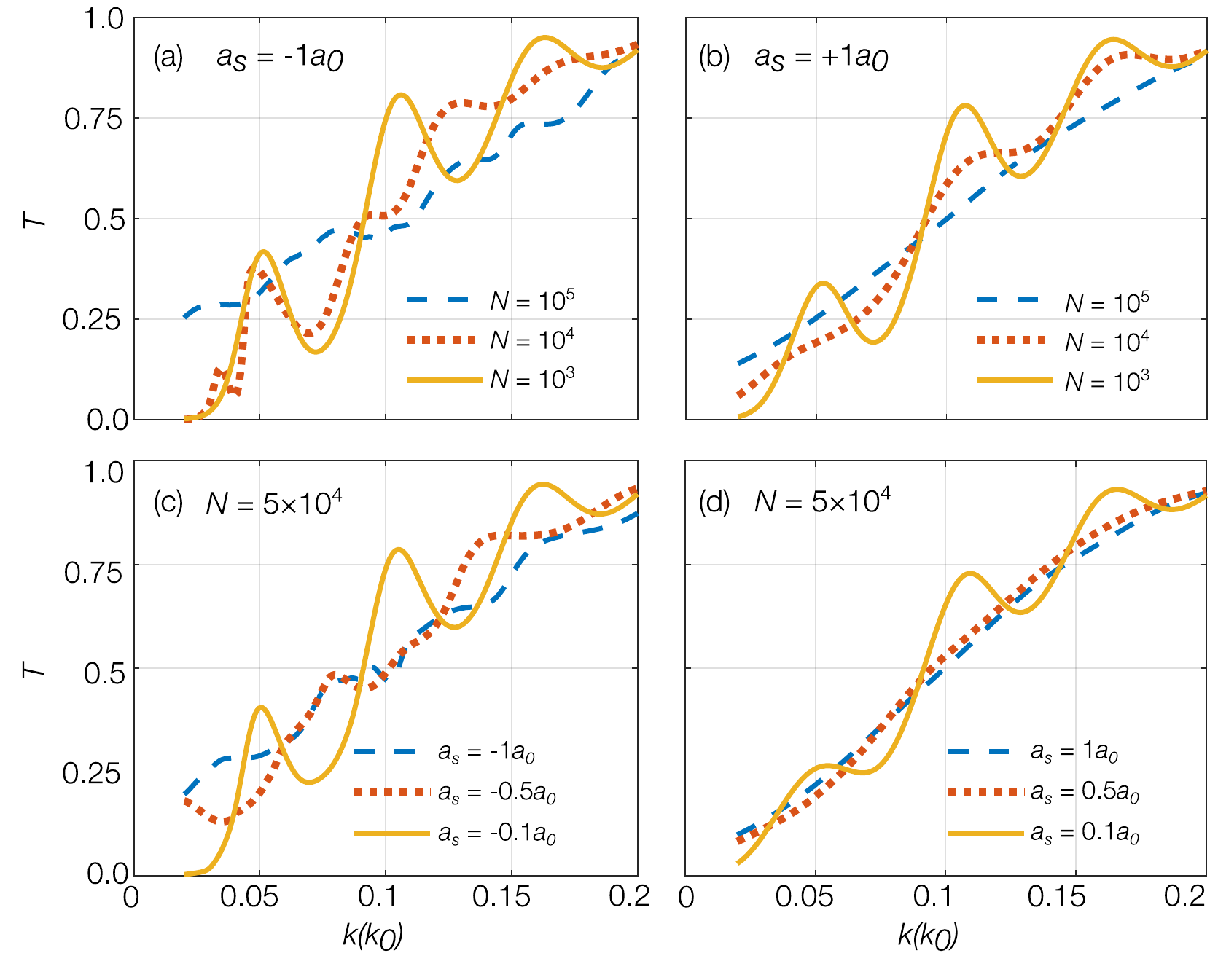}

\caption{Plots showing that reducing inter-atomic interactions improves the resonant transmission peaks. Inter-atomic interactions are reduced by reducing peak density (by reducing initial atom number) in plots (a) and (b) for $a_s=-1a_0$ and $a_s=+1a_0$, respectively. In (c) and (d), the interaction strength is reduced by reducing the magnitude of the negative and positive scattering lengths, respectively, for BECs with an initial atom number of $N=5 \times 10^4$.}
\label{fig:ReducingInteraction}
\end{figure}
In contrast to the incident plane-wave source case, where the Fabry-Perot resonance peaks occur at a maximum of $T_\text{max} = 1$~\cite{xiao_resonant_2015,uma_maheswari_quasi-bound_2009}, the transmission peaks observed here are suppressed, i.e, $T_\text{max}<1$, reducing the contrast of the resonance peaks. A previous theoretical investigation observed this behaviour for the resonance of a non-interacting CW atom laser beam~\cite{damon_reduction_2014}. The reduction in resonant transmission arises due to the finite momentum width of the source BEC. This also causes broadening of the peaks, which reduces finesse. As the peaks are suppressed due to the momentum spread of the cloud, we expect them to improve by reducing the BEC's momentum width. To investigate this further, we study the transmission profile for a range of initial cloud momentum widths. Fig.~\ref{fig:Varyingkwidthas_0} shows the height, finesse and linewidth of the first resonant peak for a non-interacting BEC with momentum width ranging from $\Delta k =2\times 10^{-4} k_0$ to 0.02 $k_0$. The resonance peaks are improved (peak height and finesse increase and linewidth decrease) by reducing the cloud momentum width, as one would expect.   

\begin{figure}[t]
\centering
\includegraphics[width=0.8\linewidth]{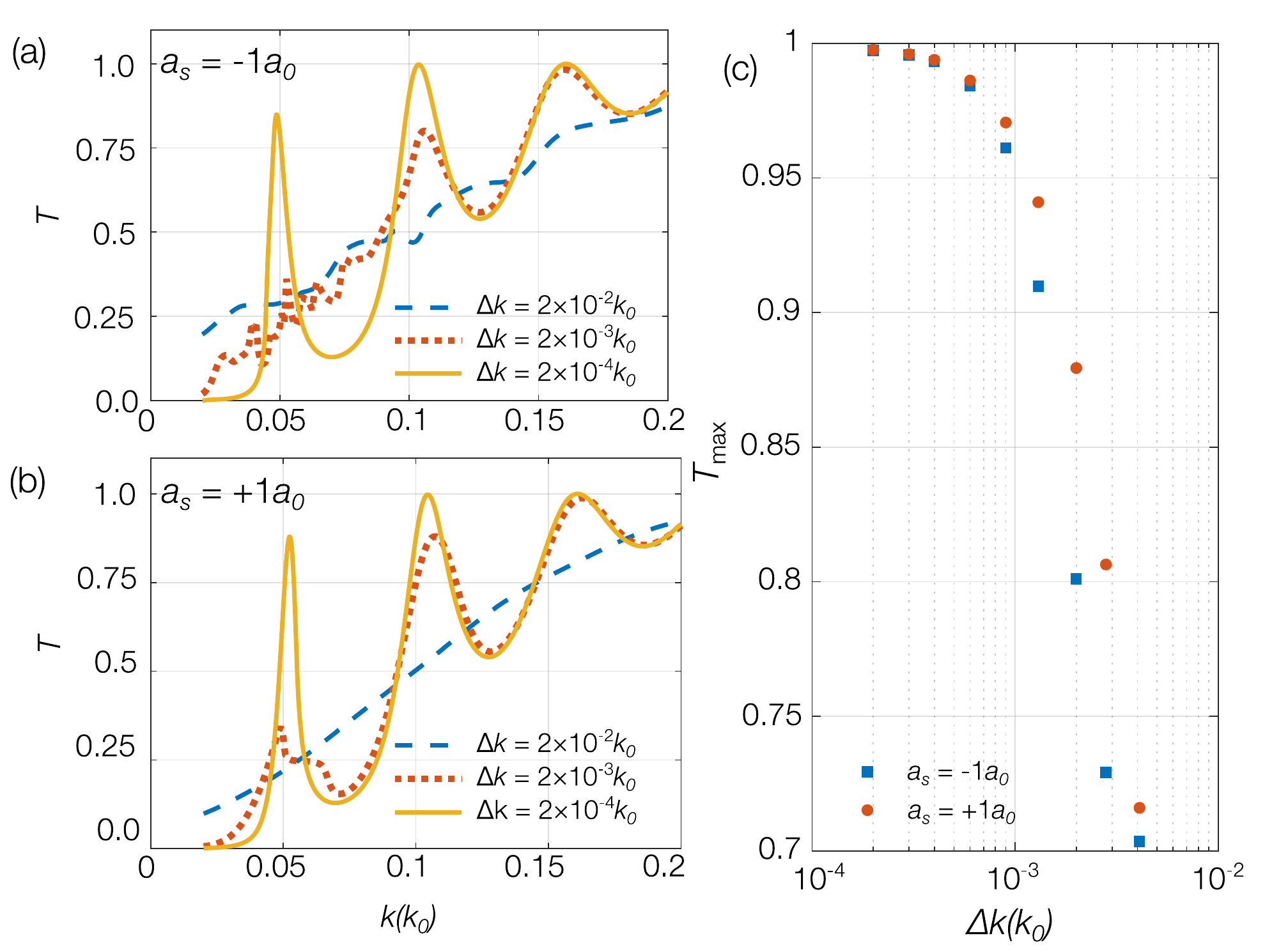}

\caption{Plots illustrating that reducing the initial momentum width of the cloud improves the transmission resonance peaks. (a) and (b) show the resonant peaks for $a_s=-1a_0$ and $a_s=+1a_0$, respectively, for three values of cloud momentum width. (c) plots the height of the second peak in the transmission spectrum ($T_\text{max}$) as a function of BEC momentum width, for scattering lengths $a_s=-1a_0$ and $a_s=+1a_0$. It shows that peak height increases with decreasing momentum width. In all plots the initial atom number is $N = 5 \times 10^4$.}
\label{fig:VaryingkWidthInteracting}
\end{figure}

\subsection*{Interacting case}
In the presence of inter-atomic interactions, the effect of three-body recombination losses become crucial. Due to the repeated reflections of the atoms between the two barriers, the density becomes very high inside the cavity, increasing the three-body recombination loss of atoms from the condensate. The overall atom loss due to three-body recombination depends non-trivially on the transmission dynamics, and so will vary with scattering length, the initial momentum kick, and spatial width of the BEC. 
Specifically, as the scattering length goes from positive to negative, the three-body recombination loss increases due to the difference in the propagation dynamics of the BEC as it approaches the barrier. Condensates with positive and negative scattering lengths undergo expanding or focussing, respectively, under free propagation~\cite{manju_quantum_2018, dekel_nonlinear_2009,carr_macroscopic_2005}. These dynamics modify the cloud density and therefore the overall three-body loss as well. The momentum kick imparted to the BEC further modifies the atom loss by changing the total interaction time. For larger momentum kicks, the atomic cloud propagates faster and spends less time in the high density region inside the cavity. This reduces the possibility of three-body recombination loss. Hence, as the average momentum of the cloud increases the loss rate decreases.  Finally, for a fixed initial atom number, increasing the spatial width of the cloud (i.e. decreasing the momentum width) decreases the overall three-body loss due to the reduction in initial density. 

The transmission resonances corresponding to clouds having $a_s = -1 a_0$ and $a_s = +1 a_0$ and a momentum width of $\Delta k=0.02 k_0$ are illustrated by the blue dashed curves in Fig.~\ref{fig:ReducingInteraction}(a) and (b). In the presence of inter-atomic interactions the peaks are either further suppressed or not well-defined, as compared to the non-interacting cloud.
This reduction in contrast is caused by the additional interaction-induced expansion of the BEC's momentum distribution, scattering-length-dependent distortions in the momentum distribution that occur during interaction with barriers~\cite{manju_quantum_2018}, and non-trivial intra-cavity dynamics due to the presence of inter-atomic interactions.

In order to mitigate the loss of contrast caused by inter-atomic interactions, we can either decrease the interactions or reduce the initial momentum width.
Both approaches effectively reduce the initial interaction energy of the cloud. Figure~\ref{fig:ReducingInteraction} shows the effects of two methods to reduce interactions, firstly by reducing initial atom number and secondly by reducing the magnitude of the scattering length. Figure~\ref{fig:ReducingInteraction}(a) and (b) show the resonance peaks for $a_s=-1a_0$ and $+1a_0$, respectively, for initial atom numbers $N=10^5, 10^4$ and $10^3$. Resonance peak contrast increases substantially by reducing initial atom number for both attractive and repulsive clouds. For comparison, Fig.~\ref{fig:ReducingInteraction}(c) and (d) illustrates the effect of reducing the magnitude of the scattering length on the resonance peaks for attractive and repulsive clouds, respectively. Similar to the effect seen from the reduction in initial atom number, reducing the magnitude of the scattering length also improves the resonant peaks for both attractive and repulsive clouds.

Although reducing the initial momentum width pushes the cloud outside of the current experimentally-realisable regime, it is important to understand the system dynamics in this narrow momentum width regime. The effect of reducing momentum width on the Fabry-Perot transmission spectrum for clouds having scattering length $a_s=-1a_0$ and $+1a_0$ are shown in Fig.~\ref{fig:VaryingkWidthInteracting}(a) and (b), respectively. Figure~\ref{fig:VaryingkWidthInteracting}(c) is produced by selecting the maximum transmission of the middle resonance. Here, the general trend towards $T_\textrm{max}=1$ is evident as the momentum width of the cloud approaches zero.

This numerical analysis shows that an atomic Fabry-Perot interferometer with a pulsed BEC source can be experimentally achieved using current technology. Since the inter-atomic interactions reduce the resonant transmission, it is ideal to use a non-interacting cloud or a very weakly-interacting cloud with small to moderate atom numbers. Additionally, as current cooling techniques improve, allowing narrower momentum width sources, it may be possible to operate a `good' quality atomic Fabry-Perot interferometer with more strongly-interacting cloud.

\section*{Conclusion}
We have compared the properties of optical and atomic Fabry-Perot interferometers. By analysing the dependence of finesse and transmission coefficient on barrier height, barrier width, cavity length, incident atomic energy and cloud momentum width, we have determined an experimentally-feasible parameter regime for observing atomic Fabry-Perot resonances. Using these parameters, we numerically simulated the transmission dynamics of a $^{85}$Rb BEC through two Gaussian barriers. The simulations showed that the Fabry-Perot resonances can be achieved for a non-interacting BEC with a momentum width around $0.02 \hbar k_0$. Due to the finite momentum width of the BEC, the transmission peaks are suppressed ($T_\textrm{max}<1$) leading to wider peaks and reduced finesse. Consequently, reducing momentum width can increase finesse and improve resonance peaks in the atomic Fabry-Perot spectrum. The introduction of inter-atomic interactions further modify and suppress the resonance peaks. We have investigated different possibilities for improving the quality of the resonant peaks of an interacting BEC, which includes reducing the interactions (by reducing initial atom number and/or the magnitude of the scattering length) and reducing the initial momentum width of the BEC. We have shown that both methods can improve the quality of resonances and we have illustrated that almost complete transmission ($T_\textrm{max}\approx 1$) is achievable for BECs having weak attractive and repulsive interactions. 
Our investigation shows that Fabry-Perot resonances can be observed only for atomic species with very low interactions or with tunable interactions, such as $^{85}$Rb. This study paves the way to experimentally realise an atomic Fabry-Perot interferometer using a weakly-interacting and non-interacting pulsed BEC, that could potentially be used for many applications including velocity filtering, accelerometry and for identifying bosonic and fermionic isotopes of an element. 

\bibliography{DoubleBarrierTunneling}

\section*{Acknowledgements}
We acknowledge useful discussions with S.A.~Haine and A.C.~White. This research was supported by funding from the Australian Research Council (ARC) Discovery Project No.~DP160104965 and was undertaken with the assistance of resources and services from the National Computational Infrastructure (NCI), which is supported by the Australian Government. SSS was supported by an ARC Discovery Early Career Researcher Award (DECRA), Project No.~DE200100495.

\section*{Author contributions statement}
The theoretical analysis and numerical simulations were carried out by PM under the supervision of KSH, PBW, NPR, and SSS. Additional numerical simulations completed during the peer-review process were performed by SSS. All authors contributed to the interpretation of results and the writing of the manuscript.

\section*{Additional information}
 \textbf{Competing interests:} The authors declare no competing interests. 

\section*{Appendix: Reduction of 3D Gross-Pitaevskii equation to non-polynomial Schr\"odinger equation} \label{subsec:1DGPE}
Consider the 3D Gross-Pitaevskii equation describing the macroscopic wave function of a BEC~\cite{Dalfovo:1999}:
\begin{align} \label{3D_GPE}
	i \hbar \frac{\partial \Psi(\mathbf{r} ,t)}{\partial t} = &\bigg[ -\frac{\hbar^2}{2 m}\nabla^2 + V_\text{ext}(\mathbf{r},t) + g|\Psi(\mathbf{r},t)|^2-i\hbar \frac{K_3}{2}|\Psi(\mathbf{r},t)|^4 \bigg]\Psi(\mathbf{r},t),
\end{align}
where $V_\text{ext}$ is an external trapping potential of the form
\begin{align}
     V_\text{ext}(\mathbf{r},t) &= \frac{1}{2} m \omega_\perp^2 (x^2 + y^2) + V(z,t) \equiv V_\perp(x,y) + V(z,t).
\end{align}
Here $m$ is the atomic mass and $\omega_\perp$ is the frequency of the radial harmonic trap. For a sufficiently tight radial confinement (large $\omega_\perp$), we can approximate the wave function as a Gaussian in the radial direction, $\phi(x,y,\sigma(z,t))$, multiplied by a 1D axial wave function $\psi(z,t)$:
\begin{equation}
    \Psi(\textbf{r},t) = \phi(x,y,\sigma(z,t)) \psi(z,t) = \frac{\exp\left[-\frac{x^2 + y^2}{2 a_\perp^2 \sigma(z,t)^2}\right]}{\sqrt{\pi}a_\perp \sigma(z,t)}  \psi(z,t), \label{wave function_ansatz}
\end{equation}
where $a_\perp = \sqrt{\hbar / (m \omega_\perp)}$, $\sigma(z,t)$ encodes the width of the radial Gaussian wave function, and $\psi(z,t)$ is normalised to the atom number, $N(t)$ (atom number can vary with time due to three-body recombination losses). 
Multiplying Eq.~(\ref{3D_GPE}) by $\phi^*(x,y,\sigma(z,t))$ and integrating over $x$ and $y$ gives,
\begin{align} \label{3DGPE_integration}
    i\hbar \frac{\partial \psi(z,t)}{\partial t}&= \Bigg[-\frac{\hbar^2}{2m}\frac{\partial^2}{\partial z^2} + V(z,t) +  g\left(\int dx dy\, |\phi(x,y,\sigma(z,t))|^4\right)|\psi(z,t)|^2 -i\hbar \frac{K_3}{2}\left(\int dx dy \, |\phi(x,y,\sigma(z,t))|^6\right)|\psi(z,t)|^4 \notag \\ 
    &+ \left(-\frac{\hbar^2}{2m} \int dx dy \, \phi^*(x,y,\sigma(z,t)) \left( \frac{\partial^2}{dx^2} + \frac{\partial^2}{dy^2}\right)\phi(x,y,\sigma(z,t))\right) \notag \\
    &+ \left(\int dx dy\, V_\perp(x,y)|\phi(x,y,\sigma(z,t))|^2\right)\Bigg] \psi(z,t).
\end{align}
Noting that
\begin{align}
    \int dx dy \,|\phi(x,y,\sigma(z,t))|^4 &=\frac{1}{2\pi a_\perp^2 \sigma(z,t)^2}, \\
    \int dx dy \,|\phi(x,y,\sigma(z,t))|^6 &=\frac{1}{3\pi^2 a_\perp^4 \sigma(z,t)^4}, \\
    -\frac{\hbar^2}{2m} \int dx dy \, \phi^*(x,y,\sigma(z,t)) \left( \frac{\partial^2}{dx^2} + \frac{\partial^2}{dy^2}\right)\phi(x,y,\sigma(z,t)) &= \frac{\hbar^2}{2 m a_\perp^2 \sigma(x,y)^2}, \\
    \int dx dy \, V_\perp(x,y)|\phi(x,y,\sigma(z,t))|^2 &= \frac{1}{2} m \omega_\perp^2 a_\perp^2 \sigma(z,t)^2,
\end{align}
we obtain the NPSE Eq.~(\ref{Eq.1DGPE}). The parameter $\sigma(z,t)$ is constrained by minimising the Gross-Pitaevskii action functional, yielding $\sigma(z,t)^2 = \sqrt{1 + 2 a_s |\psi(z,t)|^2}$~\cite{Salasnich:2002}.
\end{document}